\documentclass[onecolumn,10pt,journal]{IEEEtran}

\usepackage{amsopn}
\usepackage{amsxtra,amssymb}
\usepackage[latin1]{inputenc}
\usepackage[english]{babel}
\usepackage{graphicx}
\usepackage{hyperref}

\newcommand{\proba}{\mathbb{P}}

\usepackage{graphicx}  
\newif\ifpdf
\ifx\pdfoutput\undefined
\pdffalse 
\DeclareGraphicsExtensions{.eps} \else
\pdfoutput=1 
\pdftrue \DeclareGraphicsExtensions{.pdf,.jpg,.png} \fi

\pdfoutput=1
\begin{document}

\title{Joint Impact of Pathloss Shadowing and Fast Fading - An Outage Formula for Wireless Networks}

\author{\authorblockN{Jean-Marc Kelif$^1$, Marceau Coupechoux$^2$ \\}
\authorblockA{$^1$Orange Labs, Issy-Les-Moulineaux, France\\
jeanmarc.kelif@orange-ftgroup.com \\
$^2$TELECOM ParisTech \& CNRS LTCI, Paris, France\\
coupecho@enst.fr}}


%


\maketitle

\begin{abstract}
In this paper, we analyse the joint impact of pathloss, shadowing and fast fading on wireless networks. Taking into account the pathloss and the shadowing, we first express the SINR distribution of a mobile located at a given distance from its serving base-station (BS). The moments of this distribution are easily computed, using the Fenton-Wilkinson method, and a fluid model that considers the cellular network as a continuum of BS. Then considering the joint impact of pathloss, shadowing and fast fading, we derive an easily computable outage probability formula, for a mobile located at any distance from its serving BS. We validate our approach by comparing all results to Monte Carlo simulations performed in a traditional hexagonal network. Indeed, we establish that the results given by the formula are close to the ones given by Monte Carlo simulations.  
The proposed framework is a powerful tool to study performances of cellular networks e.g. OFDMA systems (WiMAX, LTE). 
\end{abstract}


\IEEEpeerreviewmaketitle

\section{Introduction}

The characterization of interference allows to analyze cellular networks performance. In cellular networks, an important parameter for this characterization is the SINR (signal to interference plus noise ratio) at any point of the cell. Indeed, it allows the derivation of performance evaluation (outage probabilities, capacity,...). We focus on downlink throughout this paper, since that direction is often the limited link in term of capacity. Nevertheless, our approach can easily be extended to the uplink. 

The issue of expressing outage probability in cellular networks has been extensively addressed in the literature. For this study, there are two possible assumptions: (1) considering only the shadowing effect, (2) considering both shadowing and fast fading effects. In the former case, authors mainly face the problem of expressing the distribution of the sum of log-normally random variables; several classical methods can be applied to solve this issue (see e.g. \cite{Prat00} \cite{Stu01}). In the latter case, formulas usually consist in many infinite integrals, which are uneasy to handle in a practical way (see e.g. \cite{Lin92}). In both cases, outage probability is always an explicit function of all distances from the user to interferers. 

As the need for easy-to-use formulas for outage probability is clear, approximations need to be done. Working on the uplink, \cite{Ev99} derived the distribution function of a ratio of path-losses with shadowing, which is essential for the evaluation of external interference. For that, authors approximate the hexagonal cell with a disk of same area. Authors of \cite{Veer99} assume perfect power control on the uplink, while neglecting fast fading. On the downlink, Chan and Hanly \cite{Chan01} precisely approximate the distribution of the other-cell interference. They however provide formulas that are difficult to handle in practice and do not consider fast fading. Immovilli and Merani \cite{Immo91} take into account both channel effects and make several assumptions in order to obtain simplified formulas. In particular, they approximate interference by its mean value. Outage probability is however an explicit function of all distances from receiver to every interferer.
Zorzi \cite{ZoP94}, proposes a formula available but for
packet radio networks rather than for cellular systems. Authors of \cite{PaE06}
provide some interesting characterizations and upper bounds
of the outage probability but neglects the slowly varying pathgains.
In \cite{WiP85}, authors consider both shadowing and fast fading
but assume a single interferer.

In this paper, to calculate the SINR distribution of a mobile located at any distance from its serving base station (BS), we consider the useful power he received and the interferences coming from all the BS stations of the network. We jointly take into account pathloss, shadowing and fast fading and derive a simple and easily computable outage probability formula.  We validate that formula by comparisons of the figures
obtained with it to the ones obtained by Monte Carlo simulations. At last, we rely on a recently proposed fluid model \cite{Kel06} \cite{KeA05} in order to express the outage probability as a simple analytical expression depending on the distance to the serving BS. Such an expression allows further integrations much more easily than with existing formulas. The formula can be used to analyze the coverage of a BS. 

In the next section, we derive outage probabilities, while considering first only pathloss and shadowing and then pathloss, shadowing and fast fading jointly. The computation is based on the fluid model (section \ref{fluidmodel}). In section \ref{valid}, we validate our approach and compare analytical expressions with results obtained through Monte Carlo simulations.

\section{Outage Probabilities}
We consider N interfering base station (BS), a mobile $u$ and its serving $BS_0$.

\subsection{Pathloss and Shadowing impact}
\subsubsection{Propagation} 
Considering the power $P_j$ transmitted by the BS \textit{j}, the power $p_{j,u}$ received by a mobile \textit{u} 
can be written: 
\begin{equation} \label{propag}
p_{j,u}= P_{j}Kr_{j,u}^{-\eta} Y_{j,u},
\end{equation}
where $Y_{j,u}=10^{\frac{\xi_{j,u}}{10}}$ represents the shadowing effect. The term $Y_{j,u}$ is a lognormal random variable characterizing the random variations of the received power around a mean value.  $\xi_{j,u}$ is a \textit{Normal} distributed random variable (RV), with zero mean and standard deviation, $\sigma$, comprised between 0 and 10~dB. The term $P_{j}Kr_{j,u}^{-\eta}$, where K is a constant, represents the mean value of the received power at distance $r_{j,u}$ from the transmitter $(BS_j)$. The probability density function (PDF) of this slowly varying received power is given by 
\begin{equation} \label{aqteta}
p_Y(s)= \frac{1}{a\sigma s\sqrt{\pi}}exp - \left(\frac{\ln(s)-am}{\sqrt{2}a\sigma} \right)^2
\end{equation}
where 
$a=\frac{\ln10}{10}$,
$m=\frac{1}{a}\ln(KP_jr^{-\eta}_{j,u})$ is the (logarithmic) received mean power expressed in decibels (dB), which is related to the path loss and
$ \sigma$ is the (logarithmic) standard deviation of the mean received signal due to the shadowing.

\subsubsection{SINR calculation} \label{sinrshadowing}

Considering the useful power $P_0$ transmitted by the $BS_0$, the useful power $p_{0,u}$ received by a mobile \textit{u} belonging to $BS_0$ 
can be written: 
\begin{equation} \label{propag}
p_{0,u}= P_{0}Kr_{0,u}^{-\eta} Y_{0,u}.
\end{equation}
For the sake of simplicity, we now drop index $u$ and set $r_{0,u}=r$. The interferences received by $u$ coming from all the other base stations of the network are expressed by:
\begin{equation}
p_{ext} = \sum_{j=1}^N P_j K \cdot r_{j}^{- \eta} Y_j.
\end{equation}
The SINR at user $u$ is given by:
\begin{equation}
\gamma = \frac{P_{0}Kr^{-\eta} Y_{0}}{ \sum_{j=1}^N P_j K  r_{j}^{- \eta} Y_j+N_{th}}.
\end{equation}
If now, thermal noise is negligible and BS have identical transmitting powers, the SINR can be written $\gamma = 1/Y_f$ with
\begin{equation} \label{Yf}
Y_f = \frac{\displaystyle\sum_{j=1}^N  r_j^{- \eta} Y_j}{r^{- \eta} Y_0}.
\end{equation}

\subsubsection{Distribution of $Y_f$} \label{interferencepower}
The factor $Y_f$ is defined for any mobile $u$ and it is location dependent. The numerator of this factor is a sum of log-normally distributed RV, which can be approximated by a log-normally distributed RV \cite{Stu01}. The denominator of the factor is a log-normally distributed RV. $Y_f$ can thus be approximated by a log-normal RV. 

Using the Fenton-Wilkinson \cite{Fen01} method, we can calculate the mean and standard deviation, $m_f$ and $s_f$, of factor $Y_f$, for any mobile at the distance $r$ from its serving BS, $BS_0$. We first calculate the mean and the variance of a sum of log-normal RV (appendix 1). We afterwards apply the result to the sum of $N$ log-normal identically distributed RV (appendix 2):
\begin{eqnarray}\label{mf}
m_f &=& \frac{1}{a} \ln (y_f( r, \eta) H(r, \sigma)), \\
s^2_f&=& 2(\sigma^2 - \frac{1}{a^2}\ln  H(r, \sigma)),
\end{eqnarray}
where
\begin{equation} \label{H}
H(r,\sigma) = e^{a^2 \sigma^2/2} \left( G(r,\eta)(e^{a^2 \sigma^2} -1 ) +1 \right)^{-\frac{1}{2}},
\end{equation}
\begin{equation} \label{G}
G(r, \eta) = \frac{\sum_j r_j^{-2 \eta}}{ \left( \sum_j r_j^{-\eta} \right)^2},
\end{equation}
\begin{equation} \label{feta}
y_f(r, \eta) = \frac{\sum_j r_j^{ -\eta}}{ r^{-\eta}}.
\end{equation}

From Eq.~\ref{Yf} and \ref{feta}, we notice that $y_f(r, \eta )$ represents the factor $Y_f$ \textit{without shadowing}.

\subsubsection{Outage probability}

The outage probability is defined as the probability for the SINR $\gamma$ to be lower than a threshold value $\delta$ and can be expressed as:
\begin{equation}
\proba  \big(\gamma < \delta \big) = 1-\proba  \big(p_0  > \delta (p_{ext} + N_{th} ) \big)
\end{equation}

So we have:
\begin{eqnarray} 
\proba  \left ( \gamma < \delta \right ) &=& 1-\proba  \left ( \frac{1}{\delta} >  Y_f(m_f, s_f)  \right ) \\
&=& 1-\proba  \left (10 \log_{10}(\frac{1}{\delta}) > 10 \log_{10}(Y_f) \right ) \\
&=& Q \left [ \frac{10 \log_{10}(\frac{1}{\delta})-m_f}{s_f} \right ] .
\end{eqnarray}
where $Q$ is the error function: $Q(u) = \frac{1}{2} erfc(\frac{u}{\sqrt{2}})$

The outage probability for a mobile \textbf{located at a distance r} from its serving BS, taking into account shadowing can be written:
\begin{equation} \label{CDFsha}
\proba \big( \gamma < \delta  \big) =  Q \left [ \frac{10 \log_{10}(\frac{1}{\delta})-m_f}{s_f} \right ] .
\end{equation}

\subsection{Pathloss Shadowing and Fast fading impact}
\subsubsection{Propagation} 
The power received by a mobile $u$ depends on the radio channel state and varies with time due to fading effects (shadowing and multi-path reflections). Let $P_j$ be the transmission power of a base station $j$ and $X_{j,u}$ the fast fading for mobile $u$, the power $p_{j,u}$ received by a mobile \textit{u} 
can be written: 

\begin{equation} \label{propagfading}
p_{j,u} = P_j K r_{j,u}^{- \eta} X_{j,u}Y_{j,u},
\end{equation}

where $X_{j,u}$ is a RV representing the Rayleigh fading effects, whose pdf is $p_X(x)=e^{-x}$.  

\subsubsection{SINR calculation}
The interference power received by a mobile $u$ can be written:
\begin{equation}
p_{ext,u} = \sum_{j=1}^N P_jK r_{j,u}^{- \eta} X_{j,u}Y_{j,u}.
\end{equation}
Considering that the thermal noise is negligible and that all BS transmit with the same power $P_0$, we can write (dropping index $u$):
\begin{equation}
\gamma=\frac{r^{-\eta} X_0Y_{0}}{ \sum_{j=1}^N  r_{j}^{- \eta} X_jY_j}.
\end{equation}

\subsubsection{Outage probability} 

As previously, the outage probability is expressed as:
\begin{equation}
\proba\big(\gamma < \delta \big)=1-\proba\big(r^{- \eta} X_0Y_0 > \delta (\sum_{j=1}^N r_j^{- \eta} X_jY_j  ) \big).
\end{equation}

The interference power received by a mobile $u$ due to fast fading effects varies with time. As a consequence,  the fast fading can increase or decrease the power received by $u$.  
We consider that the increase of interfering power due to fast fading coming from some base stations are compensated by decrease of interfering powers coming from other base stations. As a consequence, for the interfering power, only the slow fading effect has an impact on the SINR. So, we assume that $X_j$ can be neglected  $\forall$ j $\neq$ 0 (this assumption will be validated by simulations in the next section):
$\proba\big(r^{- \eta} X_0Y_0 > \delta (\sum_{j=1}^N r_j^{- \eta} X_jY_j  ) \big) \approx \proba\big(r^{- \eta} X_0Y_0 > \delta (\sum_{j=1}^N r_j^{- \eta} Y_j  ) \big)$.
\vspace{0.5cm}

So we have:
\begin{eqnarray} \nonumber
\proba\big(\gamma < \delta \big)&=&1-\proba\big(X_0Y_0 > \delta \frac{1}{r^{- \eta}} (\sum_{j=1}^N r_j^{- \eta} Y_j) \big) \\ \nonumber
&=&1-\proba\big(X_0(t) > \delta Y_f \big) \\ \nonumber
&=&1-\int_0^{\infty} \proba\big( x > \delta Y_f  \big) p_{X}(x) dx \big) \\ \nonumber
&=&1-\int_0^{\infty} \proba\big( 10 \log_{10}(\frac{x}{\delta}) > 10 \log_{10}(Y_f)  \big) e^{-x} dx. \nonumber
\end{eqnarray}
As a consequence, the outage probability for a mobile \textbf{located at a distance r} from its serving BS, taking into account both shadowing and fast fading can be written:
\begin{equation} \label{CDF}
\proba \big( \gamma < \delta  \big) = \int_0^{\infty} Q \left [ \frac{10 \log_{10}(\frac{x}{\delta})-m_f}{s_f} \right ] e^{-x} dx,
\end{equation}
where $Q$ is the error function: $Q(u) = \frac{1}{2} erfc(\frac{u}{\sqrt{2}})$

\subsection*{Remark }
We notice that Eq.~\ref{CDF} characterizes the cumulative density function (CDF) of $\gamma$.
That formula also allows to analyze the coverage of a BS for a given outage probability.

\section{Analytical fluid model } \label{fluidmodel}	 
To calculate the outage probability we need to express analytically the factors $m_f$ and $ s_f$. In this aim we use the fluid model approach developed in \cite{KeCoG07}. The fluid model and the traditional hexagonal model are two simplifications of the reality. As shown in \cite{KeCoG07}, none is a priori better than the other but the latter is widely used, especially for dimensioning purposes. The key modelling step of the model consists in replacing a given fixed finite number of transmitters (base stations or mobiles) by an equivalent continuum of transmitters which are distributed according to some distribution function. We consider a traffic characterized by a mobile density and a network by a base station density  $\rho_{BS}$.  For a homogeneous network, the downlink parameter $y_f$ (Eq.~\ref{feta}) only depends on the distance r between the BS and the mobile. We denote it $y_f(r)$. From \cite{KeCoG07}, we have

\begin{equation} \label{fnw}
y_f(r)=\frac{2\pi \rho _{BS}r^{\eta}}{\eta - 2} \left [ (2R_c-r)^{2-\eta}-(R_{nw}-r)^{2-\eta} \right ]. 
\end{equation}
where  $2 R_c $ represents the distance between two base stations and $R_{nw}$ the size of the network.
We notice that the factor G (Eq.~\ref{G}) can be rewritten (generalizing Eq.~\ref{fnw}, dropping r, and expressing $y$ as a function of $\eta$) as:
\begin{equation} \label{Geta}
G(\eta) = \frac{y_f(2 \eta)}{{y_f(\eta)}^2}
\end{equation}
The higher the standard deviation of shadowing, the higher the mean and standard deviation
(see hereafter Eq.~\ref{limitmf} and \ref{limitSf}). This increase is however \textbf{bounded} as we can see hereafter. In a realistic network, $\sigma$ is generally comprised between 4 and 8 dB.

The Eq.~\ref{mf} means that the effect of the environment of any mobile of a cell, on the $y_f$ factor, is \textbf{characterized by a function $H(r, \sigma)$}. This function takes into account the shadowing and a  \textbf{topological G factor} which depends on the position of the mobile and the characteristics of the network as 

\begin{itemize}
\item 
the exponential pathloss parameter $\eta$, which can vary with the topography and more generally with the geographical environment as urban or country, micro or macro cells.
\item  
the base stations positions and  number.
\end{itemize}


From Eq.~\ref{G}, we have $0<G(r, \eta)<1$ whatever  \textit{r} and $\eta$. From Eq.~\ref{mf}, we can express:

\begin{equation} \label{limitmf}
y_f(r, \eta) \leq m_f \leq \frac{y_f(r, \eta)}{{G(r,\eta)}^\frac{1}{2}}
\end{equation}
and

\begin{equation} \label{limitSf}
s^2_f \leq 2 \sigma^2 .
\end{equation}

For low standard deviations (less then 4 or 5 dB), we can easily deduce from Eq.~\ref{mf} a low dependency of the mean value of $y$ with $\sigma$, and the total standard deviation  $s_f $ is very close to $\sigma$. 
These expressions show another interesting result. We established that $0 < G < 1$. Without the topological factor G ($G=0$), we have $H= e^{a^2\sigma^2}$ so that $m_f$ increases indefinitly. If $G=1$, then $H=1$ and 
$m_f = \frac{1}{a} \ln (y_f)$. For each given value of $\eta $, characterizing a given type of cells or environment, the topological factor G thus \textit{compensates} the shadowing effects.

\section{Performance evaluation} \label{valid}

In this section, we compare the figures obtained with analytical expressions (Eq. \ref{CDF}) to those obtained by Monte Carlo simulations. We also validate our approach and show the limit of our formulas. 

\subsection{Monte Carlo simulator} \label{Simulator}

\begin{figure}[htbp]
\centering
\includegraphics[scale=0.6]{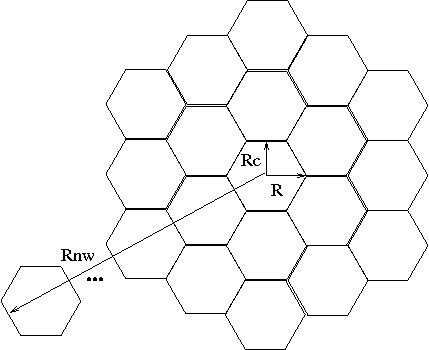}
\caption{Hexagonal network and main parameters of the study.}
\label{hexanw}
\end{figure}

The simulator assumes an homogeneous hexagonal network made of several rings around a central cell. Figure \ref{hexanw} shows an example of such a network with the main parameters involved in the study.

The simulation consists in computing the SINR for any point $u$ of the central cell. This computation can be done independently of the number of MS in the cell and of the BS output power (because noise is supposed to be negligible both in simulations and analytical study). At each snapshot, shadowing and fast fading RV are independently drawn between the MS and the serving BS and between MS and interfering BS. SINR samples at a given distance from the central BS are recorded in order to compute the outage probability. 


The validation consists in comparing results given by simulations to the ones given by the Eq.~\ref{CDF}. 

\subsection{Results}

\begin{figure}
\centering
\includegraphics[scale=0.6]{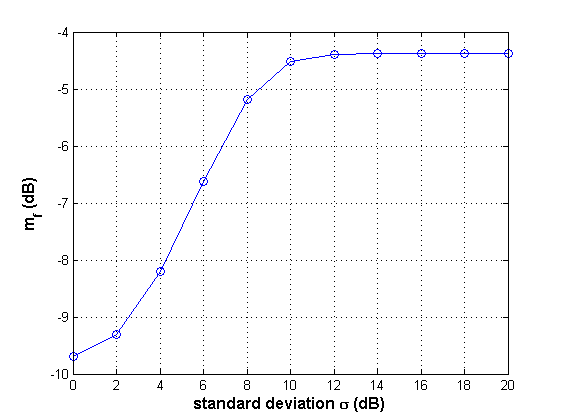}
\caption{Variations of $m_f$ [dB] with $\sigma$ [dB] for $\eta=3$.}
\label{mfvsSigma}
\end{figure}

\begin{figure}
\centering
\includegraphics[scale=0.6]{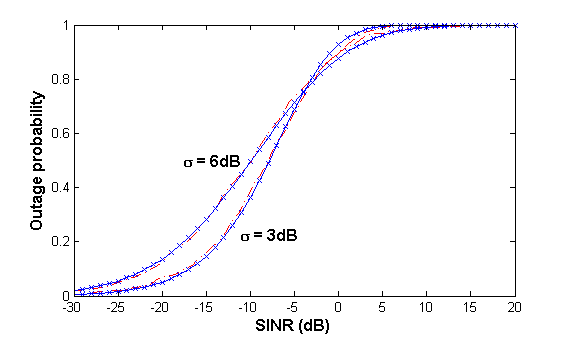}
\caption{Outage probability for a mobile located at the edge of the cell: $r=R_c$; comparison of the fluid model (solid blue curves) with simulations (dotted red curves) on an hexagonal network with $\eta=3$.}
\label{CDFSINReta3}
\end{figure}

\begin{figure}
\centering
\includegraphics[scale=0.6]{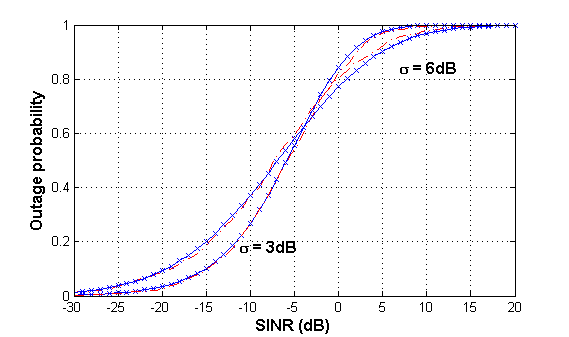}
\caption{Outage probability for a mobile located at the edge of the cell: $r=R_c$; comparison of the fluid model (solid blue curves) with simulations (dotted red curves) on an hexagonal network with $\eta=4$.}
\label{CDFSINReta4}
\end{figure}

\begin{figure}
\centering
\includegraphics[scale=0.6]{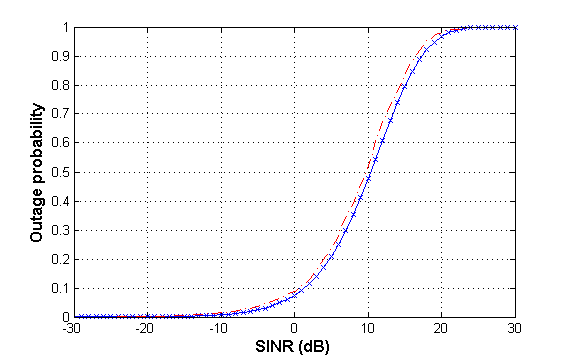}
\caption{Outage probability for a mobile located at $r=0.5 R_c$; comparison of the fluid model (solid blue curves) with simulations (dotted red curves) on an hexagonal network with $\eta=3$ and $\sigma=3dB$.}
\label{CDFSINReta3r05}
\end{figure}

\begin{figure}
\centering
\includegraphics[scale=0.6]{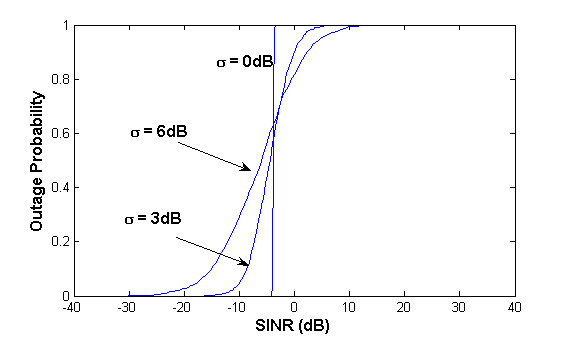}
\caption{Outage probability for a mobile located at $r= R_c$; shadowing impact (without fast fading) for $\eta=3$.}
\label{CDFSINReta4r1}
\end{figure}

Figure \ref{mfvsSigma} shows that in presence of shadowing, the value of $m_f$ increases. However, we observe that this increase reaches a maximum value for $\sigma = 12 dB$. It is due to the effect of the topology factor G which compensates the effect of shadowing. 
To validate our analytical approach, we present results for different pathloss factors $\eta$ =3 and 4, different standard deviations $\sigma =$ 3 and 6 dB, and for mobiles located at the edge of the cell $r=R_c$ and the center of the cell $r=R_c/2$. 

Figure \ref{CDFSINReta3} and \ref{CDFSINReta4} give two examples of the kind of results we are able to obtain instantaneously thanks to the formulas derived in this paper for the path-loss exponents $\eta$ =3 and 4. These curves represent the outage probability for a mobile located at the edge of the cell. Figure \ref{CDFSINReta3r05} gives outage probability considering  a path-loss exponent $\eta$ =3, for a mobile at a distance of $r=R_c/2$. 
Figures \ref{CDFSINReta3}, \ref{CDFSINReta4} and \ref{CDFSINReta3r05} show that our analytical approach gives results (solid blue curves) very close to the ones obtained with simulations (dotted red curves).

Figures \ref{CDFSINReta3}, \ref{CDFSINReta4}  \ref{CDFSINReta3r05} and \ref{CDFSINReta4r1}
show the specific impact of shadowing alone (Figure \ref{CDFSINReta4r1}) and both shadowing and fast fading (Figures \ref{CDFSINReta3}, \ref{CDFSINReta4}, \ref{CDFSINReta3r05}) on the outage probability. 
In particular the curve $\sigma = 0$~dB (Figure \ref{CDFSINReta4r1}) shows that, without shadowing and fast fading,  the SINR received at the edge of the cell reaches -4dB. Since there is no  random variations of the propagation, this value is deterministic: the outage is 0\% until a SINR of -4dB and 100\% for higher values of SINR.
When the shadowing is taken into account (but not the fast fading), we observe for $\sigma = 3dB$, a SINR of -8 dB for an outage probability of 10\%.  
And when the shadowing and the fast fading are both taken into account (Figure \ref{CDFSINReta3}), we observe for $\sigma = 3dB$, a SINR of  -16 dB for an outage probability of 10\%.    

With these results, an operator would be able to analyze the outage, and to admit or reject new connections according to each specific environment of the entering MS.
Indeed, the SINR depends on the mobile location and on the environment (Eq.~\ref{mf}), i.e. 
the shadowing,
the exponential pathloss parameter $\eta$ (which depends on different characteristics and particularly the cell dimensions and the type of environment e.g. urban or country),
the number and positions of the base stations.

\subsection{Applications}
Hereafter we present some other examples of possible applications of the analytical approach we proposed. 

\subsubsection{Coverage} The coverage of a BS can be defined as the area where the SINR received by a mobile reaches a given value with a given outage probability. Figure 4 shows, for $\eta $= 4, that at the edge of the cell (r= $R_c$), the SINR reaches the value of -15 dB with a probability of 10\%, when $\sigma = 3dB$. With the same outage probability (10\%), the SINR only reaches -20 dB when $\sigma = 6 dB$.

\subsubsection{Capacity} Since the capacity and the throughput $D$ can be expressed as a function of the SINR, for example using the Shannon expression $D \propto \log_2(1+ SINR)$, the analytical method we developed can be used to analyze this performance indicator instantaneously.

\section{Conclusion}
In this paper, we establish a simple formula of the outage probability in cellular networks, while considering pathloss, shadowing and fast fading. Taking into account pathloss and shadowing, we first express the inverse of the SINR of a mobile located at a given distance of its serving BS as a lognormal random variable. We give close form formulas of its mean $m_f$ and standard deviation $s_f$ based on the analytical fluid model. It allows us to analytically express the outage probability at a given distance of the serving BS. We then consider both pathloss, shadowing and fast fading and give an analytical expression of the outage probability at a given distance of the serving BS. The fluid model allows us to obtain formulas that only depend  on the distance to the serving BS. The analytical model that we propose is validated by comparisons with Monte Carlo simulations. The formulas derived in this paper allow to obtain performances results instantaneously. 

The proposed framework is a powerful tool to study performances of cellular networks and to design fine algorithms taking into account the distance to the serving BS, shadowing and fast fading. It can particularly easily be used to study frequency reuse schemes in OFDMA systems.

\section*{Appendix 1}
Each BS transmits a power  $P_j=P$ so the power received by a mobile is characterized by a lognormal distribution $X_j$ as $ln(X_j) \propto N(am_j,a^2\sigma^2_j)$   and we can write $m_j=\frac{1}{a}ln(P_jr^{- \eta}_j)$ (to simplify the  calculation we consider K=1).
So the total power received by a mobile is a lognormal RV X characterized by its mean and variance $ ln(X) \propto N(am,a^2\sigma^2_t)$   and we can write:
$am= \ln \left( \sum_{j=0,j\neq b}^{N} e^{(\ln P_j- \eta ln r_j+\frac{a^2 \sigma^2_j}{2})} \right) - \frac{a^2 \sigma^2_t}{2} $
so we have  since $P_j=P$ whatever the base station j, considering identical $\sigma_j=\sigma$:
\begin{equation} 
am=(\ln P+ \frac{a^2 \sigma^2}{2}) + \ln(\sum_{j=0,j\neq b}^{N}e^{- \eta \ln r_j}) - \frac{a^2 \sigma^2_t}{2} 
\end{equation}
We can express the mean interference power $ \overline{I_{ext}}$ received by a mobile as:
\begin{equation} 
ln(\overline{I_{ext}})=(\ln P+ \frac{a^2 \sigma^2}{2}) + \ln(\sum_{j=0,j\neq b}^{N} r_j^{- \eta}) - \frac{a^2 \sigma^2_t}{2}
\end{equation}
The variance $a^2 \sigma^2_t$ of the sum of interferences is written as
\begin{equation} 
a^2 \sigma^2_t =\ln \left( \frac{ \sum_j e^{(2am_j+a^2 \sigma^2)(e^{a^2 \sigma^2}-1)}}{( \sum_j exp^{am_j+ \frac{a^2 \sigma^2}{2}})^2} +1 \right) 
\end{equation}
 
Introducing  
\begin{equation} 
G(\eta) = \frac{\sum_{j=0,j\neq b}^{N} r_j^{- 2 \eta} }{ \left( \sum_{j=0,j\neq b}^{B} r_j^{- \eta} \right)^2}   
\end{equation}

the mean value of the total interference received by a mobile is given by 
\begin{equation} 
\overline{I_{ext}} =P \sum_{j=0,j\neq b}^{N} r_j^{- \eta} e^{\frac{a^2 \sigma^2}{2}} \left( ( e^{a^2 \sigma^2}-1 ) G(\eta)+1 \right)^{-1/2} 
\end{equation}
and
\begin{equation} 
a^2 \sigma^2_t =\ln \left( (e^{a^2 \sigma^2}-1) G(\eta)+1 \right) + \ln \left(e^{a^2 \sigma^2} \right)
\end{equation}

\section*{Appendix 2}
We denote $I_{int}$ the power received by a mobile coming from its serving BS.
Since the ratio of two lognormal RV's is also a lognormal RV, we can write 
the following mean and logarithmic variance:
 
\begin{equation} 
m_f =\frac{\overline{I_{ext}}}{\overline{I_{int}}}
\end{equation}
and thus, considering that: $P_b=P$, we can write, dropping the index b:
\begin{equation} 
m_f = \frac{\sum_j r_j^{- \eta} }{r^{- \eta}}  e^{ \frac{a^2 \sigma^2}{2}} \left( (e^{a^2 \sigma^2}-1) G( \eta)+1 \right)^{-1/2} 
\end{equation}
and finally, denoting:
\begin{equation} 
H(\sigma)= e^{ \frac{a^2 \sigma^2}{2} } \left( (e^{a^2 \sigma^2}-1) G( \eta)+1 \right)^{-1/2} 
\end{equation}
we have
\begin{equation} 
m_f = y_f(\eta)H(\sigma) 
\end{equation}

In dB, we can express 
\begin{equation} 
{m_f}_{dB} = \frac{1}{a} \ln (y_f(\eta) H(\sigma)) 
\end{equation}

In a analogue analysis, the standard deviation is given by $a^2 s^2_f = a^2 \sigma^2_t +a^2 \sigma^2$ so we have
\begin{equation} 
a^2 s^2_f = 2(a^2 \sigma^2-ln(H(\sigma)))  
\end{equation}



%

\end{document}